\documentclass[aoas,preprint]{imsart}

\RequirePackage[OT1]{fontenc}
\RequirePackage{amsthm,amsmath, amssymb,relsize, graphicx, graphics,booktabs, array}
\RequirePackage[numbers]{natbib}
\RequirePackage[colorlinks,citecolor=blue,urlcolor=blue]{hyperref}

\startlocaldefs
\numberwithin{equation}{section}
\theoremstyle{plain}

\endlocaldefs

\def\ca{\citeauthor}
\def\cy{\citeyear}

\begin{document}

\begin{frontmatter}
\title{Change Point Analysis of Histone Modifications Reveals Epigenetic Blocks Linking to Physical Domains}

\begin{aug}
\author{\fnms{Mengjie} \snm{Chen}\ead[label=e1]{mengjie.chen@yale.edu}},
\author{\fnms{Haifan} \snm{Lin}\thanksref{t2}\ead[label=e2]{haifan.lin@yale.edu}}
\and
\author{\fnms{Hongyu} \snm{Zhao}\thanksref{t1, t3}\ead[label=e3]{hongyu.zhao@yale.edu}
}

\thankstext{t1}{Corresponding author}
\thankstext{t2}{ Supported by a National Institutes of Health grant DP1 OD006825}
\thankstext{t3}{Supported by a National Institutes of Health grant GM59507 and a National Science Foundation grant DMS 1106738}
\runauthor{M. Chen et al.}

\affiliation{Yale University} 

\address{Program of Computational Biology and Bioinformatics\\
Yale University\\
New Haven, CT 06520\\
\printead{e1}\\
\phantom{E-mail:\ }}

\address{Yale Stem Cell Center\\
Yale School of Medicine\\
New Haven, CT 06520\\
\printead{e2}\\
\phantom{E-mail:\ }}

\address{Department of Biostatistics\\
Yale School of Public Health\\
New Haven, CT 06520\\
\printead{e3}
\phantom{E-mail:\ }}
\end{aug}

\begin{abstract}
Histone modification is a vital epigenetic mechanism for transcriptional control in eukaryotes. High-throughput techniques have enabled whole-genome analysis of histone modifications in recent years. However, most studies assume one combination of histone modification invariantly translates to one transcriptional output regardless of local chromatin environment. In this study we hypothesize that, the genome is organized into local domains that manifest similar enrichment pattern of histone modification, which leads to orchestrated regulation of expression of genes with relevant biological functions. We propose a multivariate Bayesian Change Point (BCP) model to segment the \textit{Drosophila melanogaster} genome into consecutive blocks on the basis of combinatorial patterns of histone marks. By modeling the sparse distribution of histone marks across the chromosome with a zero-inflated Gaussian mixture, our partitions capture local BLOCKs that manifest relatively homogeneous enrichment pattern of histone modifications. We further characterized BLOCKs by their transcription levels, distribution of genes, degree of co-regulation and GO enrichment. Our results demonstrate that these BLOCKs, although inferred merely from histone modifications, reveal strong relevance with physical domains, which suggest their important roles in chromatin organization and coordinated gene regulation.\end{abstract}


\end{frontmatter}

\section{Introduction}
Epigenetics refers to the study of heritable changes affecting gene expression and other phenotypes that occur without a change in DNA sequence. Epigenetic mechanisms, including chromatin remodeling, histone modification, DNA methylation and binding of non-histone proteins, provide a fundamental level of transcriptional control. Extensive studies on histone modifications have led to the ``histone code" hypothesis that histone modifications do not occur in isolation but rather in a combinatorial manner to provide ``ON" or ``OFF" signature for transcriptional events (\ca{epigenetics}, \cy{epigenetics}). 

Genome-wide studies using high-throughput technologies such as chromatin immunoprecipitation (ChIP) followed by microarray analysis (ChIP on chip) or deep sequencing (ChIP-seq) have begun to decipher the ``histone code'' at the genome-wide scale. Currently, a common approach to assess chromatin states using these data is a multivariate Hidden Markov Model (HMM) introduced by \ca{hmm:naturebiotech} (\cy{hmm:naturebiotech}), which has been employed in several modENCODE and ENCODE project publications (\ca{functional:science} \cy{functional:science}, \ca{comprehensive:nature} \cy{comprehensive:nature}, \ca{heterochromatin:gr} \cy{heterochromatin:gr}, \ca{signature:gr} \cy{signature:gr}). This model associates each 200bp genomic window with a particular state, generating a chromatin-centric annotation. However, a pre-defined number of states needs to be specified in HMMs and it is difficult to justify and interpret a particular choice. Different studies trying to balance resolution and interpretability based on different criteria often led to different numbers of states, both between different organisms (\ca{hmm:naturebiotech} \cy{hmm:naturebiotech}, \ca{functional:science} \cy{functional:science}) and within the same organism (\ca{protein:cell} \cy{protein:cell}, \ca{functional:science} \cy{functional:science}). Moreover, HMM summarizes chromatin information in a vector of ``emission" probabilities associated with each chromatin state and a vector of ``transition" probabilities with which different chromatin states occur in spatial relationship of each other (\ca{hmm:naturebiotech} \cy{hmm:naturebiotech}). These settings assume the homogeneity of hidden states and their transitions across the genome. However, since histone modifications are outcomes of interplay with local environment, the assumption of spatial homogeneity may not hold at the genome level. 

To address the limitations in the HMM-based approaches, we propose an alternative approach to examining combinatorial histone marks at coarse scales. We hypothesize that the genome is organized into local blocks that display regionalized histone signatures. Those blocks may have important roles in orchestrated regulation of expression of genes with relevant biological functions. We note that our approach does not require a pre-defined number of possible states and it identifies local patterns without the assumption on spatial homogeneity. 

To computationally infer these blocks, we propose a multivariate Bayesian Change Point (BCP) model which is capable of incorporating both local and global information. The BCP model was first proposed by \ca{hartigan:2} (\cy{hartigan:2, hartigan:1}) to describe a process where the observations can be considered to arise from a series of contiguous blocks, with distributional parameters different across blocks. One of the inferential goals is to identify the change points separating contiguous blocks. By ``assuming probability of any partition is proportional to a product of prior cohesions, one for each block in the partition, and that given the blocks the parameters in different blocks have independent prior distributions"  (\ca{hartigan:2} \cy{hartigan:2, hartigan:1}), a fully Bayesian approach can be adopted to detect change points from a sequence of observations. \ca{hartigan:2} (\cy{hartigan:2}) considered in detail the case where the observations $X_{1}$, ... ,$X_{n}$ are independent and normally distributed given the sequence of parameters $\mu_{l}$ with $X_{i} \sim N(\mu_{l}, \sigma^2)$ where the observations from the same block $l$ have the same $\mu_l$. This method has been used by \ca{Emerson:1} (\cy{Emerson:1}) to segment microarray data. However, this model cannot be directly applied to infer histone modification blocks because observed modification data do not follow normal distributions. This is due to the fact that histone modifications are usually observed at a small proportion of the genome locations with remaining of the input signal being (or near) zero (Figure S1). Moreover, individual histone modifications may have spatially shifted pattern because of physical constraint on occupancy. To address these unique features, here we report a new multivariate BCP model through the introduction of a zero-inflated Gaussian mixture distribution, to partition the genome into blocks where each block is relatively homogeneous with respect to histone marks.

\subsection{Outline of the Paper}

We organized the paper as following. In Section \ref{sec:method}, we present the methodological details of the BCP model with a mixture prior and an MCMC algorithm to infer the posterior probability. Section \ref{sec:simu} presents results from simulation studies. In Section \ref{sec:app}, we describe a systematic change point analysis of the \textit{D. \textit{melanogaster}} genome with a compendium of histone marks in S2 cells with modENCODE data.The identified chromosomal blocks are called as BLOCKs in the rest of this article. In the following Section, we present two sets of exploratory analysis, with Section \ref{sec:simu} studying BLOCKs' relationship with physical domains and Section \ref{sec:simu} investigating the functional relevance of BLOCKs. In Section \ref{sec:simu}, we compare our results with HMM. We conclude the paper with a summary and discussion in Section \ref{sec:disc}.

\subsection{Notations}

We denote the density function of $N(\mu,\sigma^2)$ by $\phi(\cdot|\mu,\sigma)$, and denote the density function of $\text{Beta}(a,b)$ by $\psi(\cdot|a,b)$. The Dirac function $\delta$ indicates the point mass at $0$. For a set $S$, $\#S$ is the cardinality of $S$. For a random variable $X$, $\{X=1\}$ is the indicator function taking value $1$ if $X=1$ and taking value $0$ if $X\neq 1$. The indicator function $\{X=0\}$ is defined in the same way. The set $\{i+1,i+2,...,j\}$ with integers $i<j$ is denoted by $(i:j)$. The function $f(\cdot|\cdot)$ is a generic notation for conditional density when the distribution is clear in the context.

\section{Method} \label{sec:method}

\subsection{A BCP model for block identification}

The observation we have is a  $M\times n$ data matrix $\mathbb{X}=(\mathbf{X}_1,...,\mathbf{X}_{M})^T$, where each $\mathbf{X}_m$ for $m=1,...,M$ is a modification mark with length $n$. We first describe the likelihood of each $\mathbf{X}_m$ and then combine them together. For notational simplicity, we suppress the subscript and write $\mathbf{X}$ instead of $\mathbf{X}_m$. 

Let $\mathbf{X}=(X_1,...,X_n)$ be a vector with length $n$. Create another vector $\mathbf{Z}=(Z_1,...,Z_n)$ to indicate whether each $X_h$ is zero or not. That is, $Z_h=0$ if $X_h=0$, and $Z_h=1$ if $X_h\neq 0$. Note $\mathbf{Z}$ is fully determined by $\mathbf{X}$.

For the index set $\{1,...,n\}$, let $\rho$ be a partition of this set. That is $\rho=\{S_1,...,S_N\}$, with $\{1,...,n\}=\bigcup_{l=1}^N S_l$ and $S_{l_1}\bigcap S_{l_2}=\varnothing$ for all $l_1\neq l_2$. The number $N$ represents the number of blocks of $\{1,...,n\}$. For the change-point problem, each $S_l$ is a contiguous subset of $\{1,...,n\}$. That is, $S_l=(i:j)=\{i+1,...,j\}$ for some $i<j$.

\subsubsection{Likelihood}

Given the partition $\rho=\{S_1,...,S_N\}$, $X_k$ follows a mixture distribution $X_k\sim (1-\lambda)N(\mu_l,\sigma^2)+\lambda\delta$, for $k\in S_l$ and each $l=1,...,N$. The parameter $\mu_l$ is block-specific, while $\sigma$ and $\lambda$ are shared among different blocks. The parameter $\lambda$ describes how likely $X_k$ is zero. Thus, given $(\rho,\mu_1,...,\mu_N,\lambda,\sigma)$, the likelihood of $(\mathbf{X},\mathbf{Z})$ can be fully specified. That is,
\begin{equation}
L(\mathbf{X},\mathbf{Z}|\rho,\mu_1,...,\mu_N,\lambda,\sigma)=\prod_{l=1}^N f(X_{S_l},Z_{S_l}|\mu_l,\lambda,\sigma), \label{eq:like}
\end{equation}
where for each $l$ with $S_l=\{i+1,...,j\}$, 
\begin{eqnarray}
&& f(X_{S_l},Z_{S_l}|\mu_l,\lambda,\sigma) \\
\label{eq:likerep} &=& (1-\lambda)^{\#\{k\in S_l:Z_k=1\}}\lambda^{\#\{k\in S_l:Z_k=0\}}\prod_{\{k\in S_l:Z_k=1\}}\phi(X_k|\mu_l,\sigma),
\end{eqnarray}
where $X_{S_l}=(X_{i+1},...,X_{j})$ and $Z_{S_l}=(Z_{i+1},...,Z_j)$.

\subsubsection{Prior}

We proceed to specify the prior distribution on the parameters $(\rho,\mu_1,...,\mu_N,\lambda,\sigma)$.
\begin{eqnarray}
\label{prior:par} \rho &\sim& \prod_{l=1}^N c(S_l), \\
\label{prior:mu} \mu_l &\sim& N\Big(\mu_0,\sigma_0^2d_l^{-1}\Big)\quad\text{for each $l$ with $S_l=\{i+1,...,j\}$}, \\
&& \text{and} \quad   d_l = \#\{k\in S_l:Z_k=1\}, \nonumber \\ 
\label{prior:lambda} \lambda &\sim& \text{Beta}(a,b). 
\end{eqnarray}
The prior (\ref{prior:par}) on the partition $\rho$ is called product partition model, which was originally described in \ca{hartigan:1} (\cy{hartigan:1}). The quantity $c(S_l)$ is called cohesion. In this paper, $c(S_{l})$ is defined to be $c_{(i:j)}=(1-p)^{j-i-1}p$ when $j <n$ and $c_{(i:j)}=(1-p)^{j-i-1}$ when $j=n$, where $0\leq p \leq 1$ and $S_l = \{ i+1, ... , j\}$ as mentioned before. This specification implies that the sequence of change points forms a discrete renewal process with inter-arrival times identically geometrically distributed. The priors (\ref{prior:mu}) and (\ref{prior:lambda}) are conjugate priors with respect to the likelihood. The prior on the variance $\sigma^2$ will be jointly specified with the hyper-parameters.

To pursue a fully Bayesian approach, we put priors on the hyper-parameters $(p,\mu_0,\sigma_0)$ in (\ref{prior:par}) and  (\ref{prior:mu}). Define $w=\frac{\sigma^2}{\sigma^2+\sigma_0^2}$. We jointly specify the priors on the hyper-parameters  together with the prior on $\sigma^2$.
\begin{eqnarray}
\label{prior:mu0} \mu_0 &\sim& 1,\quad-\infty<\mu_0<\infty \\
\label{prior:sigma} \sigma^2 &\sim& \frac{1}{\sigma^2}, \quad 0\leq\sigma^2<\infty, \\
\label{prior:w} w&\sim&\frac{1}{w_0},\quad 0\leq w\leq w_0, \\
\label{prior:p} p&\sim&\frac{1}{p_0},\quad 0\leq p\leq p_0.
\end{eqnarray}
The priors (\ref{prior:mu0}), (\ref{prior:w}) and (\ref{prior:p}) are uniform priors. They reflect our ignorance of knowledge. The prior (\ref{prior:sigma}) can  be viewed as a uniform distribution on the logarithmic scale. Notice (\ref{prior:mu0}) and (\ref{prior:sigma}) are improper priors. This will not cause problem in view of our sampling procedure described later. 

\subsubsection{Posterior}

Our goal here is to find the posterior distribution of the partition, which is $f(\rho|\mathbb{X},\mathbb{Z})$. According to Bayes formula,
\begin{equation}
f(\rho|\mathbb{X},\mathbb{Z})=\frac{\prod_{m=1}^M f(\mathbf{X}_m,\mathbf{Z}_m|\rho)f(\rho)}{\int \prod_{m=1}^M f(\mathbf{X}_m,\mathbf{Z}_m|\rho)f(\rho)d\rho}. \label{eq:post}
\end{equation}
Since the denominator of (\ref{eq:post}) is complicated, we need to use MCMC to sample from the posterior by
\begin{equation}
f(\rho|\mathbb{X},\mathbb{Z})\propto \prod_{m=1}^M f(\mathbf{X}_m,\mathbf{Z}_m|\rho)f(\rho).
\end{equation}
The conditional density $f(\mathbf{X},\mathbf{Z}|\rho)$ is by integrating out the likelihood function (\ref{eq:like}) using the prior of $(\mu_1,...,\mu_N,\lambda,\sigma)$ specified in (\ref{prior:mu}), (\ref{prior:lambda}), (\ref{prior:mu0}), (\ref{prior:sigma}) and (\ref{prior:w}). The prior $f(\rho)$ is by integrating out $f(\rho|p)$ specified in (\ref{prior:par}) with respect to (\ref{prior:p}). We first find $f(\rho)$.
\begin{eqnarray}
\nonumber f(\rho) &=& \frac{1}{p_0}\int_0^{p_0}f(\rho|p)dp = \frac{1}{p_0}\int_0^{p_0}\Bigg(\prod_{l=1}^N c(S_l)\Bigg)dp \\
\nonumber &=& \frac{1}{p_0}\int_0^{p_0}\Bigg(\prod_{S_l=\{i+1,...,j\}}c_{ij}\Bigg)dp \\
\label{eq:priorpar} &=& \frac{1}{p_0}\int_0^{p_0} p^{N-1}(1-p)^{n-N}dp.
\end{eqnarray}
Then, we continue to find $f(\mathbf{X},\mathbf{Z}|\rho)$. We first integrate out $(\mu_1,...,\mu_N,\lambda)$ in (\ref{eq:like}) using (\ref{prior:mu}) and (\ref{prior:lambda}). Remember $\psi(\lambda,b)$ is the density of $\text{Beta}(a,b)$. Using (\ref{eq:likerep}) as the representation of (\ref{eq:like}), we have
{\small  \begin{eqnarray}
\label{density1} && f(\mathbf{X},\mathbf{Z}|\rho,\mu_0,w,\sigma) \\
\nonumber &=& \prod_{k=1}^N\int \prod_{\{k\in S_l:Z_k=1\}} \phi(X_k|\mu_l,\sigma)\phi\big(\mu_l|\mu_0,\sigma_0 d_l^{-1/2}\big)d\mu_l \\
\nonumber&& \times\int_0^1\prod_{k=1}^N (1-\lambda)^{\#\{k\in S_l:Z_k=1\}}\lambda^{\#\{k\in S_l:Z_k=0\}}\psi(\lambda|a,b)d\lambda\\
\nonumber&=& \prod_{\{(i:j)=S_l \in \rho\}} A \times  (2\pi\sigma^2)^\frac{-T}{2} w^{\frac{N}{2}} \exp \Bigg( -\frac{1}{2\sigma^2} \Big(W+wB+wT(\mu_{0}-\bar{X}_{T} )^2\Big) \Bigg),
\end{eqnarray} }
where 
\begin{eqnarray}
\nonumber T &=& \sum_{k=1}^{n}\{Z_{k}=1\}\\
\nonumber \bar{X}_{T} &=&  T^{-1}\sum_{k=1}^n X_k\\
\nonumber \bar{X}_{(i:j),Z_{k}} &=& \frac{1}{\#\{Z_{k}=1\}}\sum_{\{k:Z_{k}=1, i < k \leq j\}}X_k\\
\nonumber W &=&\sum_{\{(i:j)=S_l \in \rho\}} \sum_{\{k:Z_{k}=1, i < k \leq j\}}(X_{k}-\bar{X}_{(i:j),Z_{k}} )^2\\
\nonumber B &=& \sum_{\{(i:j)=S_l \in \rho\}}\#\{Z_{k}=1: i < k \leq j\} (\bar{X}_{(i:j),Z_{k}}- \bar{X}_{T})^2\\
\label{A_equation} A &=& \prod_{\{(i:j)=S_l \in \rho\}} \mathsmaller{ \frac{ \Gamma( a+\#\{Z_{k}=1: i <  k \leq j\}) \Gamma( b +\#\{Z_{k}=0: i <  k \leq j\}) } { \Gamma(a+b+j-i) }} .
\end{eqnarray}
Next, we integrate out $(\mu_0,w,\sigma)$ in (\ref{density1}) using priors (\ref{prior:mu0}), (\ref{prior:sigma}) and (\ref{prior:w}).
\begin{eqnarray}
&& f(\mathbf{X},\mathbf{Z}|\rho) \\
&=& \frac{1}{w_0}\int_0^{w_0}\int \sigma^{-2}\int f(\mathbf{X},\mathbf{Z}|\rho,\mu_0,w,\sigma)d\mu_0d(\sigma^2)dw \\
\label{density2} &\propto& A \int_{0}^{w_{0}} \frac{w^{\frac{N-1}{2}}} { [W+wB] ^\frac{T+1}{2}} dw.
\end{eqnarray}
To model multiple histone marks, $\mathbf{X}_1,...,\mathbf{X}_M$ are independent vectors given the same block structure $\rho$. As has been calculated in (\ref{density2}), for each $m$,
\begin{equation}
f(\mathbf{X}_m,\mathbf{Z}_m|\rho) \propto A_m \int_{0}^{w_{0}} \frac{w^{\frac{N-1}{2}}} { [W_{m}+wB_{m}] ^\frac{T_{m}+1}{2}} dw,\label{density3}
\end{equation}
where $a_{m}$, $b_{m}$, $W_{m}$, $B_{m}$ $T_{m}$ and $A_{m}$ are values for the $m$-th sequence as $a$, $b$, $W$, $B$, $T$ and $A$ defined above. $\mathbf{Z}_{m}$ are indicators determined by $\mathbf{X}_{m}$ and $Z_{k,m}$ is the $k$-th element in $\mathbf{Z}_{m}$. Combining (\ref{eq:priorpar}) and (\ref{density3}), we have
 \begin{eqnarray}
\label{density4} f( \rho|\mathbb{X}, \mathbb{Z}) &\propto&  \Bigg( \frac{1}{p_0}\int_0^{p_{0}}p^{N-1}(1-p)^{n-N} dp\Bigg)^{M}  \\
\nonumber && \times \prod_{m=1}^{M} A_m \times \prod_{m=1}^{M}\int_{0}^{w_{0}} \frac{w^{\frac{N-1}{2}}} { [W_{m}+wB_{m}] ^\frac{T_{m}+1}{2}} dw
 \end{eqnarray}
Although an exact implementation of this model is tractable, the calculations are $O(n^{3})$. It is prohibitive to evaluate the posterior probability when n is large. We have implemented an MCMC approximation that greatly facilitates the estimation.

\subsection{MCMC algorithm for BCP model inference}

 Following \ca{hartigan:1} (\cy{hartigan:1}), for a partition $\rho$ induced by $\mathbf{U} = (U_{1}, ...,U_{n})$, where $U_{i} = 1$ indicates a change point at position $i+1$, the odds ratio for the conditional probability of a change point at the position $i+1$ is:
 \begin{eqnarray*}
&&  \frac{P(U_{i}=1|\mathbb{X}, \mathbb{Z}, U_{j}, j \neq i)} {P(U_{i}=0|\mathbb{X}, \mathbb{Z}, U_{j}, j \neq i)}  \\
&=& \frac{ \Big( \int_0^{p_{0}}p^{N}(1-p)^{n-N-1} dp\Big)^{M}  \times \prod_{m=1}^{M} A_m^1  \int_{0}^{w_{0}} \frac{w^{\frac{N-1}{2}}} { [W^{1}_{m}+wB^{1}_{m}] ^\frac{T_{m}+1}{2}} dw }{ \Big( \int_0^{p_{0}}p^{N-1}(1-p)^{n-N} dp\Big)^{M}  \times \prod_{m=1}^{M} A_m^0 \int_{0}^{w_{0}} \frac{w^{\frac{N-2}{2}}} { [W^{0}_{m}+wB^{0}_{m}] ^\frac{T_{m}+1}{2}} dw }
 \end{eqnarray*}
where $W^{0}_{m}$, $B^{0}_{m}$, $W^{1}_{m}$ and $B^{1}_{m}$ are the within and between block sums of squares obtained for the $m$-th sequence when $U_{i} = 0$ and $U_{i} = 1$ respectively, $A_m^0$ and $A_m^1$ is the values of (\ref{A_equation}) obtained for the $m$-th sequence when $U_{i} = 0$ and $U_{i} = 1$ respectively. The result is a direct consequence of (\ref{density4}).

We then approximate these integrals by incomplete beta integrals as:
\begin{eqnarray*}
&& \mathlarger{\frac{P(U_{i}=1|\mathbb{X}, \mathbb{Z}, U_{j}, j \neq i)} {P(U_{i}=0|\mathbb{X}, \mathbb{Z}, U_{j}, j \neq i)}}\\
&=&\mathlarger{\prod_{m=1}^{M}\Bigg(\Big(\frac{W^{1}_{m}}{B^{1}_{m}}\Big)^{\frac{1}{2}} \Big(\frac{W^{0}_{m}}{W^{1}_{m}}\Big)^{\frac{S_{m}-N+1}{2}}\Big(\frac{B^{0}_{m}}{B^{1}_{m}}\Big)^{\frac{N+2}{2}}\Bigg) } \\
&&  \mathlarger{\times \frac{  \prod_{m=1}^{M}\int_{0}^{\frac{B_{m}^{1}w_{0}/W^{1}_{m}}{1+B_{m}^{1}w_{0}/W^{1}_{m}}} x^{(N-1)/2} (1-x)^{T_{m}-N-2} dx}{\prod_{m=1}^{M}\int_{0}^{\frac{B_{m}^{0}w_{0}/W^{0}_{m}}{1+B_{m}^{0}w_{0}/W^{0}_{m}}} x^{(N-2)/2} (1-x)^{T_{m}-N-3} dx}} \\
&& \mathlarger{\times \frac{ \Big(\int_0^{p_{0}}p^{N}(1-p)^{n-N-1} dp\Big)^{M} \times \prod_{m=1}^{M} A_m^1 } {  \Big(\int_0^{p_{0}}p^{N-1}(1-p)^{n-N} dp\Big)^{M}\times   \prod_{m=1}^{M}  A_m^0 }}.
\end{eqnarray*}

 We initialize $U_{i}$ to 0 for all $i < n$, with $U_{n} = 1$. Then we update $U_{i}$ by passes through data. 500 passes were used in block identification.  
 
 \section{Simulation studies} \label{sec:simu}
 
\begin{figure}
\begin{center}
\includegraphics[width=5in]{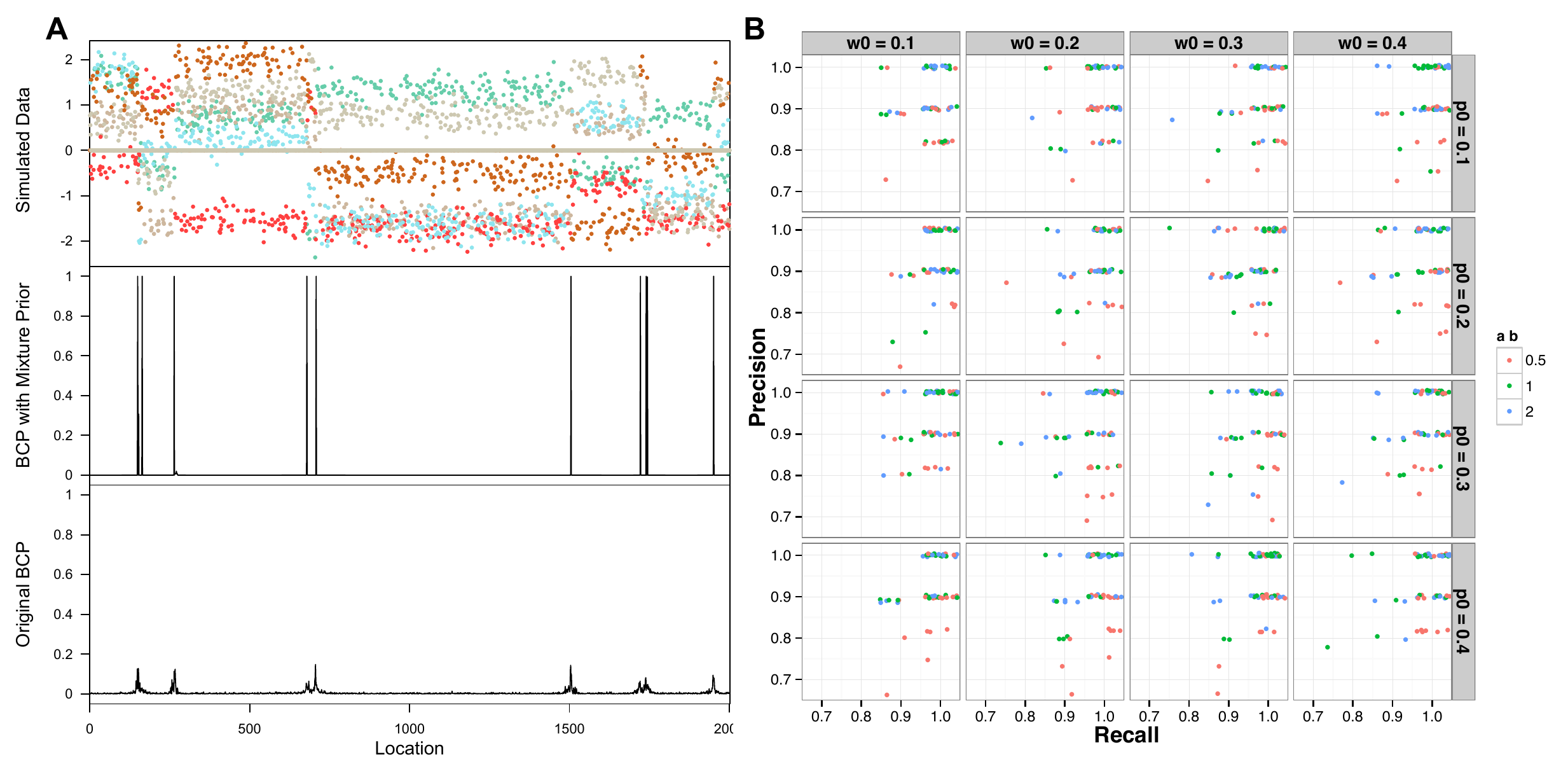}
\end{center}
\caption{Simulation results. A. One example of simulated datasets with posterior probabilities inferred from proposed BCP model with $p_{0}=0.1$, $w_{0}=0.1$, $a_{m}=b_{m}=0.5$ and from original BCP model using function bcp() in R package bcp. B. Jitter plot for precision and recall rates of BCP model with 48 different sets of hyper-parameters on 20 simulated datasets. }
\label{fig:1}
\end{figure}

First we used simulated data to study the performance of the proposed method. The simulation assumed that there were 10 blocks and six histone modification marks were observed at each one of the 2000 locations in the genome. The lengths of the 10 blocks were ranging from 10 to 1500 (In simulation 1 shown in Figure \ref{fig:1}, the lengths are $152, 10, 102, 416, 27, 799, 217, 22, 206$ and $49$). We use $X_{(i:j), m}$ to denote the observed signal within a block from $({i+1})$-th to ${j}$-th location for the $m$-th mark. We assumed that each component of the $X_{(i:j), m}$ followed a mixture distribution of $0.2*N(\mu_{(i:j), m}, 1)+ 0.8*\delta$ where $\mu_{(i:j), m}$ was a random draw from $U(-2, 2)$. These settings are based on the empirical observation that for a specific histone mark, on average $\sim$20\% of the genome display binding peaks with the intensities ranging from -2 to 2 for the normalized data. To apply our method, we need to specify the values of the hyper-parameters $p$, $w$, $a_{m}$ and $b_{m}$. In the simulation, we investigated the sensitivity of the results to the specifications of these parameter values by considering a range of values, with  $p=(0.1, 0.2, 0.3, 0.4)$, $w=(0.1,0.2,0.3,0.4)$, and $(a_{m}, b_{m})=\{(1,1),(2,2),(0.5,0.5)\}$. As a result, we considered a total of 48 specifications for $(p_{0},w_{0},a_{m},b_{m})$. We simulated 20 data sets. For each simulated data set, we ran 48 MCMC chains with each chain using one of the 48 different hyperparameters described above. Change points were inferred to be those locations in the genome that had a posterior probability larger than 0.8 (The results were similar under different cutoff values). 

We then checked the precision and recall rates based on the true and inferred change points from the simulated data.  The precision rate is defined as TP/(TP +FP), and the recall rate is TP/(TP +FN), where TP is the number of true positives (predicted block boundaries that are true), FP is the number of false positives (predicted boundaries that are not true), and FN is the number of false negatives (undiscovered true block boundaries). In our assessment, if the inferred change point was 3 units or less from one of the true change points, this inference was considered a true positive.  As shown in Figure \ref{fig:1}B, the overall posteriors are insensitive to the specified values of the hyperparameters $p_{0}$, $a_{m}$ , $b_{m}$, however the best average precision and recall rates were obtained when $p_{0}=0.1$ and $w_{0}=0.1$. We thus used $p_{0}=0.1$, $w_{0}=0.1$, $a_{m}=b_{m}=0.5$ in later analysis. Simulation studies also showed that the proposed method is capable of identifying large blocks expanded over 1000 position as well as small blocks of size around 10 (Figure \ref{fig:1}). Moreover, the ability of identifying zero-inflated blocks is significantly boosted by the introduction of the mixture priors (Figure \ref{fig:1}).

\section{Application to modENCODE epigenome Data} \label{sec:app}

All data used in this analysis were generated by the modENCODE project (Table \ref{tab1}), including pre-processed regions of significant enrichment for 18 histone modifications in S2 and 10 in BG3 cells from experiment ``Genomic Distributions of Histone Modifications", mapped reads for S2 cell transcriptome from ``Paired End RNA-Seq of \textit{Drosophila} Cell Lines" and pre-processed SAM files with multimapped reads for 9 different developmental stages from experiment ``Developmental Stage Timecourse Transcriptional Profiling with RNA-Seq".  To identify blocks from histone modifications and then characterize them, the \textit{Drosophila} \textit{melanogaster} genome was first divided into 1000-bp bins, and the average enrichment level was calculated within each bin based on log2 intensity values using all histone modification and chromosomal protein binding profiles, and the average transcription level (in S2 cell and different development stages) was calculated within each bin based on counts of short reads taking into account individual replicates. 

\begin{table}[!ht]
\tiny
\caption{Overview of modENCODE data that were used in this study}

\begin{tabular}{>{\centering}m{3.3cm}>{\centering}p{1.2cm}>{\centering}m{2.5cm}>{\centering}m{4.5cm}}
 \hline
\bf{modENCODE Experiment} & \bf{Method} & \bf{Cell Line or Tissue Type} & \bf{Sample} 
\tabularnewline
 \hline
Genomic Distributions of Histone Modifications & ChIP-chip & S2-DRSC, ML-DmBG3-c2  & H3K18ac, H3K23ac, H3K27Ac, H3K27Me3, H3K36me1, H3K36me3, H3K4Me3, H3K4me1, H3K4me2, H3K79Me2, H3K79Me1, H3K9ac, H3K9me2, H3K9me3, H4AcTetra, H4K16ac, H4K5ac, H4K8ac 
\tabularnewline
 \hline
Transcriptional profiling of Drosophila cell lines & RNA-seq & S2-DRSC &  
\tabularnewline
 \hline
Developmental Stage Timecourse Transcriptional Profiling& RNA-seq &  Embryo 10-12h, White Pre-pupae 24h, Larvae L1, Adult Female Eclosure 1d&  
\tabularnewline
 \hline
 \end{tabular}
\begin{flushleft}
\end{flushleft}
\label{tab1}
\end{table}

\subsection{Identification of chromatin blocks based on histone modifications} \label{subsec:identification}
BCP model was applied to the genome-wide occupancy profiles for 18 different histone methylation and acetylation marks in S2 cells of \textit{Drosophila} \textit{melanogaster} from the modENCODE project. For each histone mark, we calculated the average enrichment level at non-overlapping 1kb resolution based on modENCODE called enrichment peaks. We then inferred the block structure of each chromosome separately based on the enrichment of multiple histone marks. Change points with posterior probability greater than 0.75 were defined as block boundaries. Because chromosome X is distinguished by high level of H4K16ac in combination with H3K36me3 from other chromosomes (\ca{comprehensive:nature} \cy{comprehensive:nature}), we applied our model to autosomes only. 

\begin{figure}
\begin{center}
\includegraphics[width=4.5in]{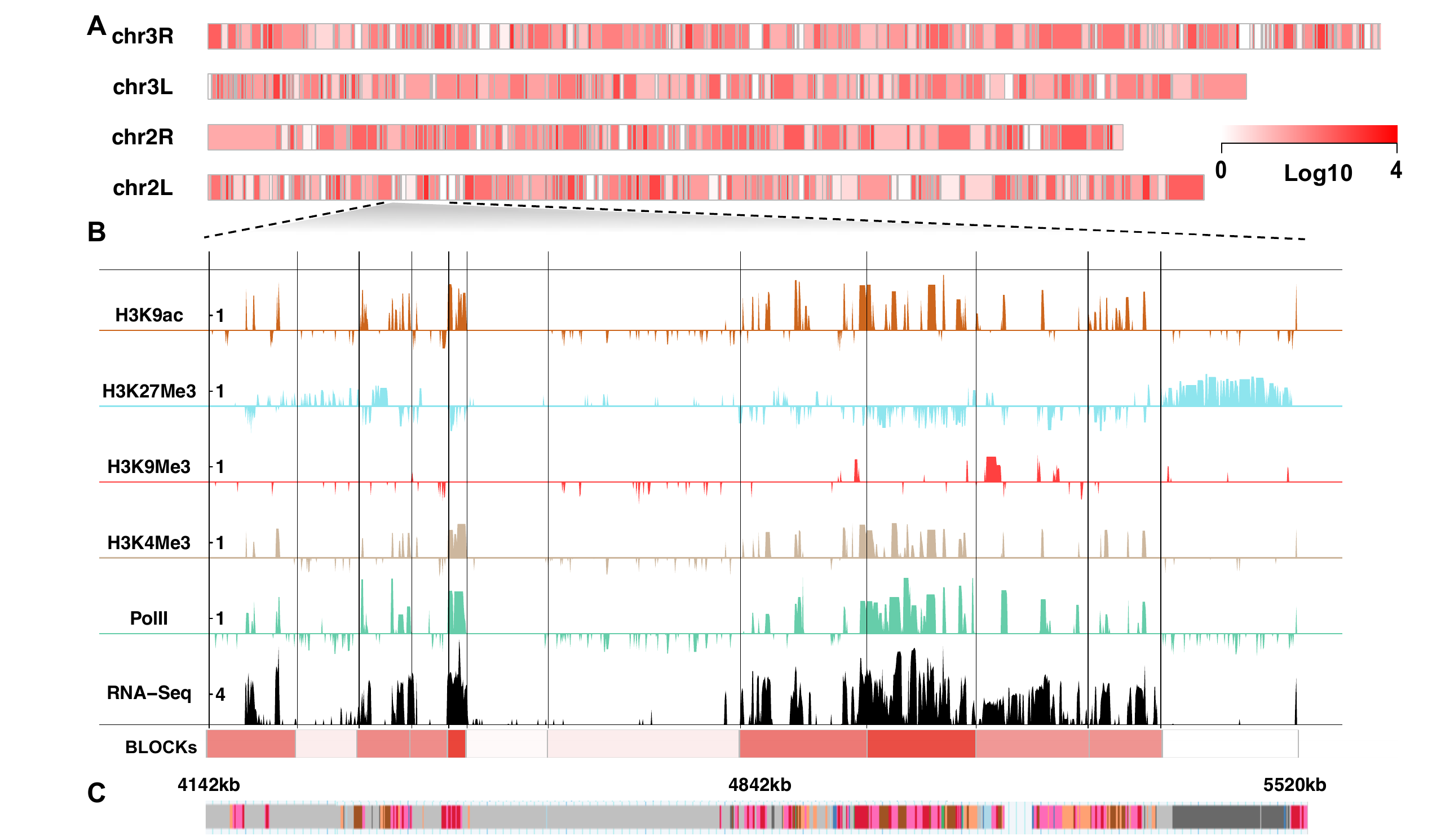}
\end{center}
\caption{BLOCKs inferred from multiple histone modifications in \textit{Drosophila melanogaster} S2 cell. A. Overview of the BLOCKs in S2 cells with average transcriptional levels shown in gradient. B. Example of BLOCK characterization at a specific locus on chromosome 2L. BLOCK boundaries are shown as solid black lines. The enrichment levels of several chromatin signatures are shown at 1kb resolution, including transcription activation marks H3K4m3 and H3K9ac, transcription repression marks H3K9me3 and H3K27Me3. PolII and RNA-seq counts at log10 scale are shown as a reference of transcriptional activity. C. ``Chromatin states" annotation from \ca{comprehensive:nature} (\cy{comprehensive:nature}).}
\label{fig:4}
\end{figure}

\begin{figure}
\begin{center}
\includegraphics[width=4.5in]{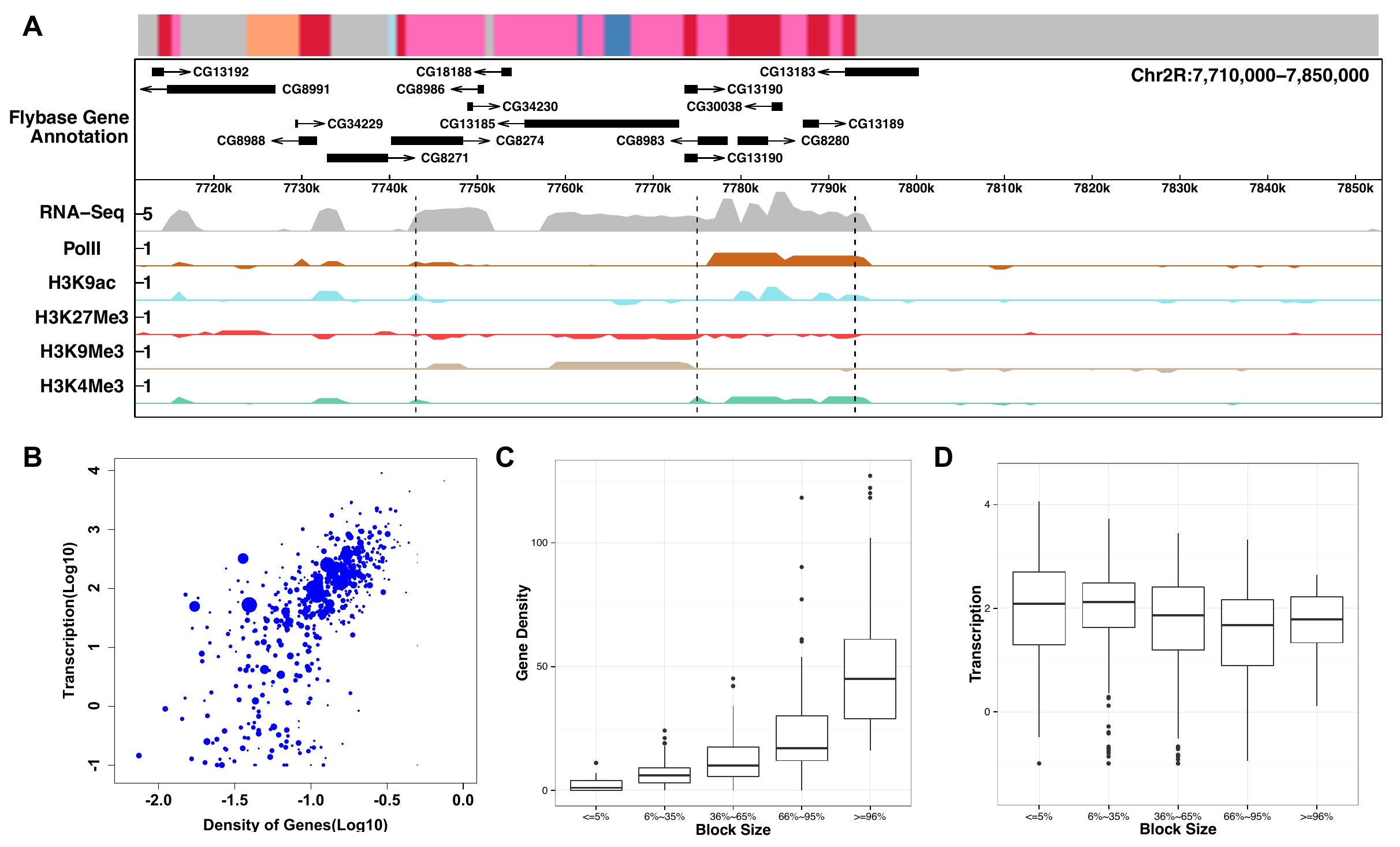}
\end{center}
\caption{BLOCKs characterization. A. A locus on chromosome 2R with four BLOCKs display diverse sizes, gene density and transcription activity (corresponded ``chromatin states" annotation from \ca{comprehensive:nature} (\cy{comprehensive:nature}) shown on the top). B. Transcription activity vs. gene density with block size shown in gradient. C. Box plot for gene density on five block size quantiles. D. Box plot for transcription activity on five block size quantiles.}
\label{fig:5}
\end{figure}

A total of 728 blocks were inferred from chromosomes 2L, 2R, 3L and 3R, with 90\% of the blocks ranging in size from 25kb to 341kb, with a median of 99kb (called as BLOCKs, Table S1). We observed that BLOCKs captured the combinatorial pattern of histone modifications and reflected local transcriptional activities. We use chr2L:4142-5520kb as an example to illustrate this (Figure \ref{fig:4}). For simplicity, we only show the enrichment levels of several chromatin signatures including transcription activation marks H3K4m3 and H3K9ac, and transcription repression marks H3K9me3 and H3K27me3 (see Figure \ref{fig:alignment} for an example of all marks). PolII enrichment and RNA-seq counts at log10 scale are shown as a reference of transcriptional activity.  Compared with ``chromatin states" annotation for non-overlapping 200bp windows in the genome (\ca{comprehensive:nature} \cy{comprehensive:nature}) (Figure \ref{fig:4}C), BLOCKs depict the genome as local domains at a larger scale. We divided BLOCKs into five quantiles based on their sizes: $\leq5\%$, $6\%\sim35\%$, $36\%\sim65\%$, $66\%\sim95\%$, $\geq96\%$ and looked into the transcription activity distributions for each group (Figure \ref{fig:5}E). Transcription activities do not show a systematic bias as a function of block size.

\subsection{BLOCK boundaries are potentially physical domain boundaries}  \label{subsec:boundary}
A recent published high-resolution chromosomal contact map on \textit{Drosophila} embryonic nuclei (\ca{cell:physical} \cy{cell:physical}) showed that the entire genome is linearly partitioned into well-demarcated physical domains. We therefore studied the link between physical domains and BLOCKs that we inferred from histone modifications. A total of 966 physical domains were identified from \textit{Drosophila} embryonic nuclei (\ca{cell:physical} \cy{cell:physical}) chromosome 2L, 2R, 3L and 3R with the sizes ranged from 10kb to 823kb and a median of 60kb. We observed strong association between physical domains and BLOCK boundaries. For example, 38\% of BLOCK boundaries are within 10kb of physical domain boundaries whereas this proportion never exceeds 26\% in 1000 randomized block partitions and 56\% of BLOCK boundaries are within 20kb of physical domain boundaries whereas this proportion never exceeds 42\% in 1000 randomized block partitions. 

\begin{figure}
\begin{center}
\includegraphics[width=5in]{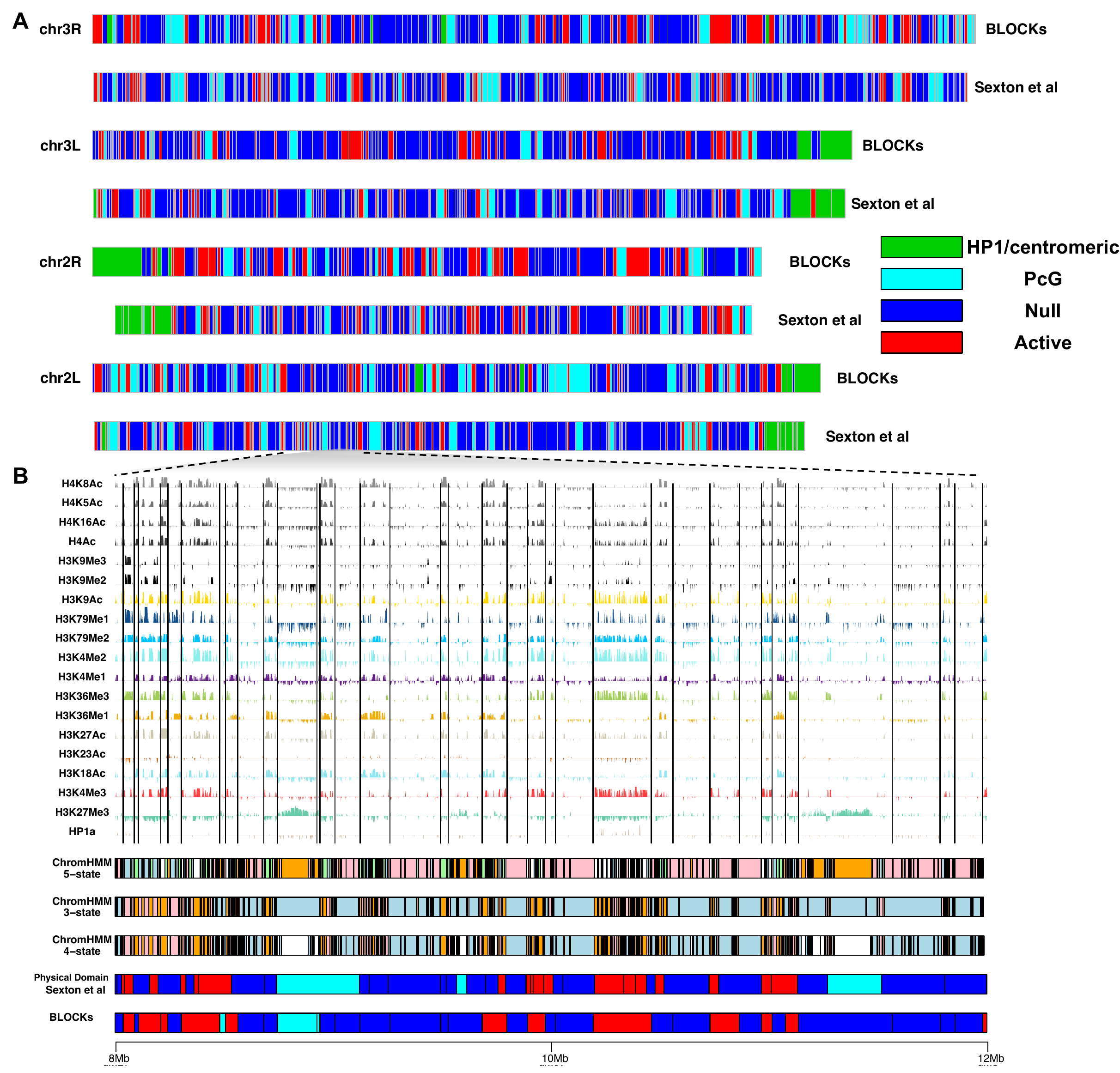}
\end{center}
\caption{A. The alignment of BLOCKs (S2 cells) with physical domains in \ca{cell:physical} (\cy{cell:physical}) (embryonic nuclei cells). B. A comparison of ChromHMM, BLOCKs and physical domains at a locus on chromosome 2L (8Mb-12Mb). BLOCK boundaries are shown as vertical gray lines.}
\label{fig:alignment}
\end{figure}

In (\ca{cell:physical} \cy{cell:physical}), the authors characterized physical domains into four epigenetic classes based on the enrichment of epigenetic marks. Out of the four classes, transcriptional ``Active" domains are associated with H3K4me3, H3K36me3, and hyperacetylation, ``PcG" domains are associated with the mark H3K27me3, ``HP1/Centromere" class is associated with HP1 and ``Null" domains are not enriched for any available marks. We explored whether BLOCKs can be aligned to the classification in (\ca{cell:physical} \cy{cell:physical}). We assigned the four classes to BLOCKs based on enrichment of H3K4me3, H3K27me3 and HP1a. For BLOCKs, ones with average intensities of HP1a greater than 1 and coverage greater than 10\% are classified as ``HP1/Centromere" domains, ones with average intensities of H3K27me3 greater than 0.5 and coverage greater than 25\% are classified as ``PcG" domains, ones with average intensities of H3K27me3 greater than 1 and coverage greater than 25\% are classified as ``Active" domains and all remaining are characterized as ``Null'' domains. Figure \ref{fig:alignment} shows the alignment between BLOCKs and physical domains with epigenetic classes. The high concordance between BLOCKs and physical domains suggests that BLOCKs bridge the link between epigenetic domains with topological domains. The difference may be introduced by techniques, data quality and cell types used in the two studies. 

Another indirect evidence for BLOCKs as physical domains is the consistency with replication timing. Replication timing refers to the order in which segments of DNA along the length of a chromosome are duplicated. Since the packaging of DNA with proteins into chromatin takes place immediately after the DNA is duplicated, replication timing reflects the order of assembly of chromatin. Recent studies suggest that late-replicating regions generically define not only a repressed but also a physically segregated nuclear compartment. Thus replication timing is a manifestation of spatial organization of the chromosome. To investigate the association of BLOCKs with replication timing, we compared BLOCKs with the meta peaks of replication origins (10kb to 285kb) from cell lines BG3, Kc and S2 analyzed by modENCODE project. We observed that 58\% of meta peaks are within 20kb of BLOCK boundaries. This statistic agrees with physical domains well since we observed that 60\% of meta peaks within 20kb of physical boundaries characterized in \ca{cell:physical} (\cy{cell:physical}). 

\subsection{Functional relevance of BLOCKS} \label{subsec:coexp}
To investigate whether BLOCKs represent domains of functional importance, we performed three different analyses. 
First, we checked whether genes within each BLOCK tended to be co-regulated using transcriptome in L1 larvae and 10-12h embryo measured by RNA-seq. A total of 11376 FlyBase genes were used in our analysis. When a gene had multiple isoforms, only the isoform with the broadest genomic occupancy was used. We defined the expression change status of each gene in L1 larvae stage (and 10-12h embryo) using expression levels in S2 cell as a reference by the following rule: genes whose expression increased by more than 2 fold but were not below 10 were categorized as ``up-regulated"; those with fold change less than 0.5 but the expression levels were not below 10 as ``down-regulated"; and others as ``no-change". To examine whether each BLOCK is enriched for genes with specific expression change patterns, we used the proportion of blocks that the most dominant pattern accounted for 50\% or more of total number of genes within that block as the test statistic. We observed the percentage of BLOCKs where the most dominant pattern accounted for more than 50\% of the genes was 61.6\% and 60.8\% for L1 larvae and 10-12h embryo, respectively, with 46.8\% of the BLOCKs showing the same pattern between the two comparisons. These observed statistics reach statistical significance by comparing with randomly permutated blocks. For physical domains in \ca{cell:physical} (\cy{cell:physical}), we observed 68\% and 65.8\% with dominant co-regulation patterns for L1 larvae and 10-12h embryo, respectively.

Second, we asked whether genes within each BLOCK tended to have similar biological functions. We tested for enrichment of Gene Ontology (GO) categories within each BLOCK by using hypergeometric test with Bonferroni correction. 52.5\% (351 out of 669 BLOCKs with more than 2 genes) were enriched for at least one GO category using a 0.05 cutoff and 1061 GO categories in total are enriched (Table S2). The observed numbers of GO enriched BLOCKs and enriched GO categories were both significantly higher than those from permutated blocks. We further asked which biological processes or functions involve genes that are significantly linearly juxtaposed. We found 90\% (108/119) of chromatin assembly or disassembly genes (GO:0006333) for \textit{Drosophila} were juxtaposed within a BLOCK located on chr2L: 21329-215856kb, with a striking p-value of $3.95\times10^{-234}$. Genes in chitin-based cuticle development (GO:0040003), structural constituent of peritrophic membrane (GO:0016490), body morphogenesis(GO:0010171), proteinaceous extracellular matrix (GO:0005578) were found significantly clustered with over 70\% genes in one BLOCK share the same function.

 Third, we reasoned if BLOCKs reflected coordinated regulation of genes with relevant biological functions, we would expect that BLOCKs enriched in developmentally specific GO categories would display large deviation in transcription across different developmental stages, while BLOCKs enriched in ``house-keeping" GO categories would display limited fluctuations. We ranked the BLOCKs based on their standard deviation of transcription across 9 different developmental stages (Table S3 and S4). BLOCKs with the top 20\% largest deviations and 20\% smallest deviations were checked for their GO enrichment respectively, and then were listed in Tables S2 and S3 in the order of statistical significance. Notably, in BLOCKs displaying most striking changes in developmental transcriptomes, we found GO categories associated with conspicuous developmental-specific biological processes or functions, specification of segmental identity, eg. heart development, structural constituent of chitin-based cuticle, positive regulation of muscle organ development, and midgut development, among others. Moreover, metabolism-related functions, such as serine-type endopeptidase activity, peptidyl-dipeptidase activity etc, display turnover across developmental transcriptomes and are among the top of our list. GO categories associated with ``house-keeping" functions, like transferase activity, aminoacylase activity, chromatin assembly, insulin receptor binding showed limited fluctuations through development. This result provides further evidence on the role of BLOCKs in coordinated regulation.

\subsection{Comparison with ChromHMM} \label{subsec:ChromHMM}
\begin{figure}
\begin{center}
\includegraphics[width=3.5in]{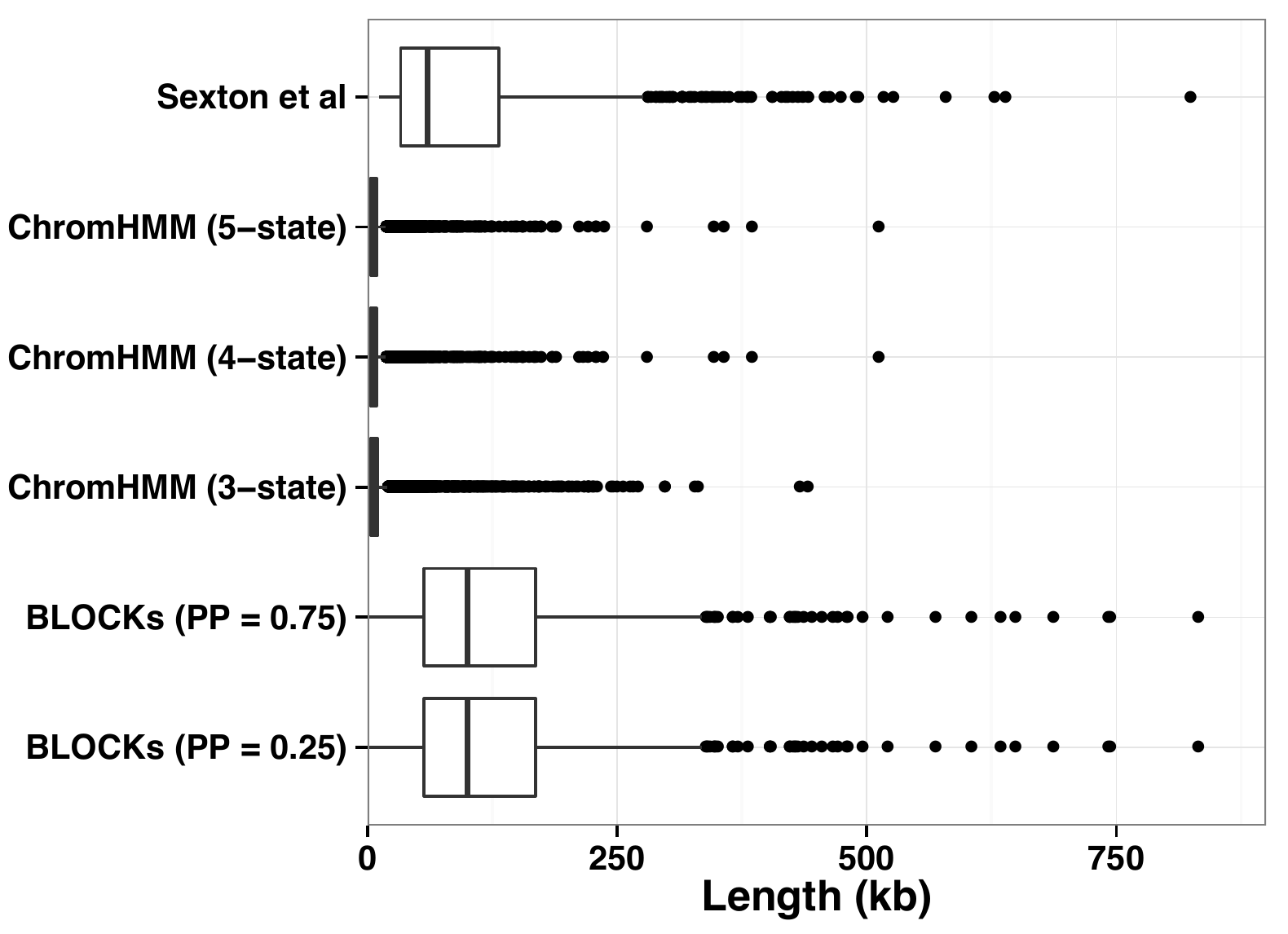}
\end{center}
\caption{Boxplot for the sizes of segments identified using different methods: physical domains in embryonic nuclei identified using High-C data (\ca{cell:physical} \cy{cell:physical}), ChromHMM with 5, 4 and 3 hidden states and BLOCKs with posterior probability greater than 0.75 and 0.25.}
\label{fig:blocksize}
\end{figure}

In this subsection, we compare the results from our method with those from a popular HMM based method, ChromHMM. We applied ChromHMM to the same dataset (18 histone modification, 1kb bins, S2 cell). The data were binarized to fit ChromHMM's requirement of input. More specifically, all intervals with intensities greater than 0 are set to 1 and remaining are set to 0. To obtain blocks at coarse levels, we explored ChromHMM models by varying the pre-specified number of hidden states (from 3 to 18). We observed that a smaller number of hidden states tended to produce blocks with larger sizes. Here we report ChromHMM models with the number of hidden states from 3 to 5. The ChromHMM model with 3 hidden states generates 12517 segments, the model with 4 hidden states generates 9157 segments, and the model with 5 hidden states generates 12444 segments.  For each ChromHMM model, the sizes of segments range from 2kb (5\% quantile) to 26kb (95\% quantile) and a median of 5kb. The distributions of sizes of segments from ChromHMM models and BLOCKs are visualized in Figure \ref{fig:blocksize}. Therefore, we think that, compared to BLOCKs, the HMM models are not able to characterize the more global histone modification patterns. Therefore, we think that, compared to BLOCKs, the Hidden Markov models are not able to characterize the more global histone modification patterns. 

\subsection{How robust is the result?}
\begin{figure}
\begin{center}
\includegraphics[width=4.5in]{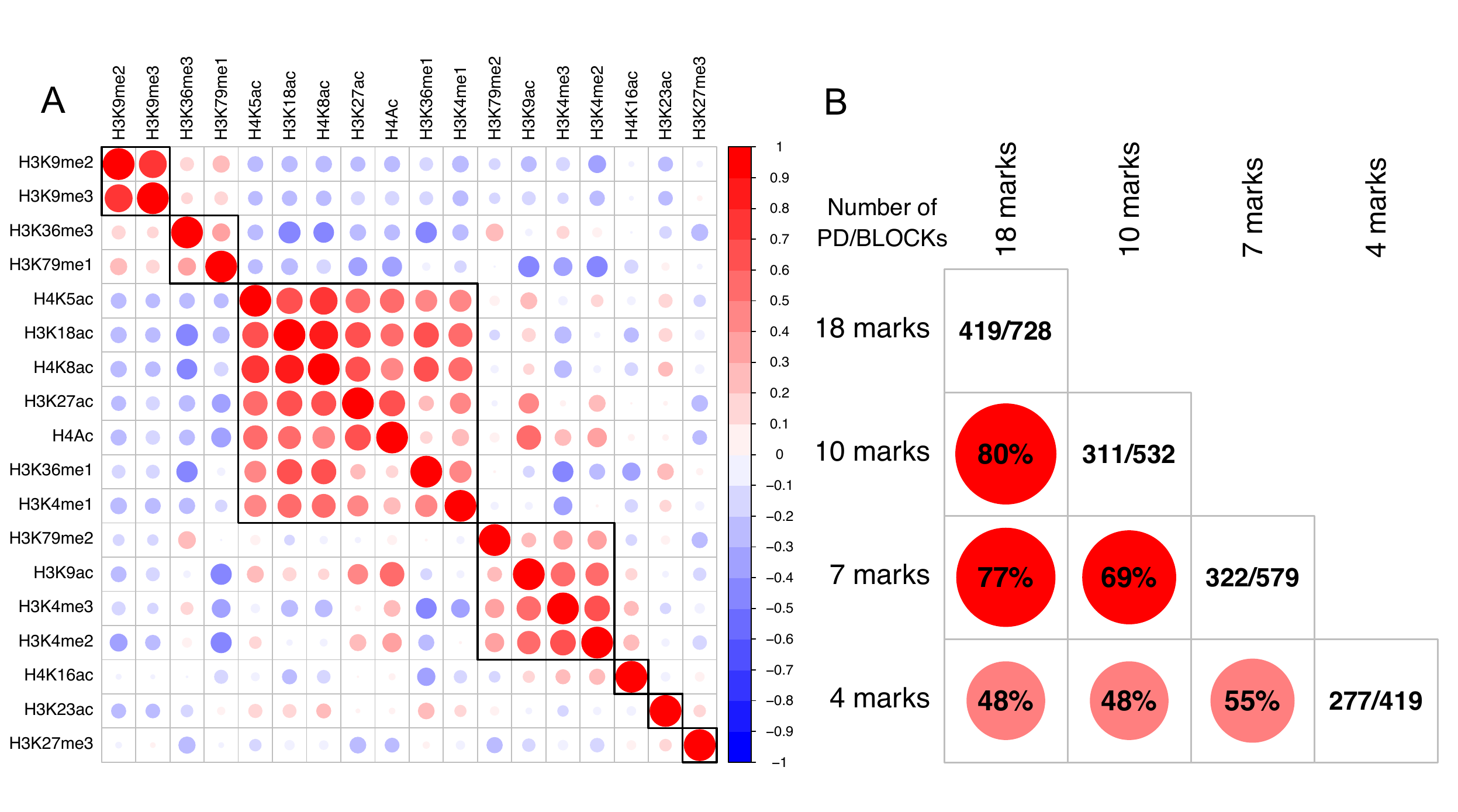}
\end{center}
\caption{A. Genome-wide correlation plot for 18 histone marks in S2 cells. The marks are ordered based on the result of hierarchical clustering. B. Comparison of models with different input histone marks. Each of the off-diagonal element is the percentage of boundaries (within 20kb) shared by any pair of the models. The diagonal element is the number of boundaries shared with physical boundaries in \ca{cell:physical} (\cy{cell:physical}) (short as PD) / the number of segments detected for each model. }
\label{fig:correlation}
\end{figure}

The BCP model used in this paper assumes that different histone marks are independent. However, some histone marks, such as H3K4me3 and H3K4me2, are highly correlated with each other. Moreover, it is known that there exists redundancy and exclusivity between the active and repressive marks. To further explore how the input histone marks will affect the result, we performed the change point analysis with the input of 4 marks, 7 marks and 10 marks, respectively. The marks for each model were selected based on their correlation across the entire genome. As shown in Figure \ref{fig:correlation}A,  there are mainly 7 groups of marks based on their correlation patterns: the first group consists of H3K9me2 and H3K9me2; the second group is featured by H3K36me3 and H3K79me1; the third group consists of H4K5ac, H3K18ac, H4K8ac, H3K27ac, H4Ac, H3K36me1 and H3K4me1; the fourth group is featured by H3K79me2, H3K9ac, H3K4me3 and H3K4me2; where as three separate groups are formed by H4K16ac, H3K23ac and H3K27me3, respectively. For the 7 marks model, we selected one mark from each of the 7 groups with the input marks as H3K18ac, H3K23ac, H3K27Me3, H3K36me3, H3K4Me3, H3K9me2, and H4K16ac. For the 10 marks model, we further introduced H4, H3K79Me2, and H3K9ac into the 7 marks model. For the 4 marks model, we excluded H3K18ac, H3K36me3, and H4K16ac from the 7 marks model. The 4 marks, 7 marks and 10 marks models identified 419, 579 and 532 blocks, respectively. We observed high consistency between these results and reported BLOCKs obtained with 18 marks, for example, 80\% of boundaries from the 10 marks model are within 20kb of BLOCK boundaries and 77\% of boundaries from the 7 marks model are within 20kb of BLOCK boundaries (see Figure \ref{fig:correlation}B for other comparisons).  

To investigate how the posterior probability cutoff would affect the characterization of BLOCKs, we varied the threshold and checked the distribution of the sizes. The results were rather stable under different cut-off values. When the cut-off value was set as low as 0.25, 40 new boundaries were added, leading to a total of 769 blocks. Although newly introduced boundaries were all almost within 10kb distance of physical domains, this number is still less than the number of physical domains identified in \ca{cell:physical} (\cy{cell:physical}).

\section{Discussion} \label{sec:disc}
\subsection{Methodological comparisons}

Our BCP model was developed with a different purpose compared to existing methods for analyzing combinatorial histone modifications. For example, ChromaSig (\ca{ChromaSig} \cy{ChromaSig}) was designed to uncover potential regulatory elements through searching for genome-wide frequently occurring chromatin signatures. Spatial clustering (\ca{spatialclustering} \cy{spatialclustering}) identified novel patterns of local co-occurrence among histone modifications by imposing a spatial K-clustering solution on HMM. Segway (\ca{segway} \cy{segway}) based on Dynamic Bayesian Networks, achieved a breakthrough in precision and resolution in finding known elements and handling of missing data compared to HMM-based approaches. The most recent method of this kind, ChAT (\ca{chat:nar} \cy{chat:nar}), extends the capabilities of chromatin signatures characterization through an inherent statistical criterion for classification. All these methods tried to detect chromatin signatures associated with a variety of small functional elements. To the best of our knowledge, our model is the first effort to examine histone marks at coarse scales although no explicit constraint has been put on block size. By separately modeling zero and non-zero signals, our model is able to capture the local enrichment patterns of vastly different sizes implicitly, superior than the existing \textit{ad hoc} merging strategy (\ca{chat:nar} \cy{chat:nar}). 

BCP differs substantially from several previously described studies to subdivide the genome at ``domain-level". 
\ca{organization:plos} (\cy{organization:plos}) reported a study to identify nested chromatin domain structure through a statistical test of each chromatin component. Their chromatin domains are specific for each component or factor whereas our approach captures domain with combinatorial pattern of multiple factors. \ca{encode:domain} (\cy{encode:domain}) used a simple two-state HMM to segment the ENCODE regions into active and repressed domains based on multiple tracks of functional genomic data, including activating and repressive histone modifications, RNA output, and DNA replication timing. By using wavelet smoothing, their method focuses on a single scale at a time (\ca{dnase:bioinformatics} \cy{dnase:bioinformatics}). In contrast, our analysis focuses on histone modifications only and simultaneously captures enrichment patterns over different scales. BCP is most similar to a four-state CPM model proposed to characterize chromatin accessibility based on tiled microarray DNaseI sensitivity data only (\ca{dnase:bioinformatics} \cy{dnase:bioinformatics}). Both methods formulate the segmentation of genome into a change point detection problem. However, these two methods differ in several respects. First, CPM is still a hidden-state model with transition probabilities imposed on segments other than equal-sized bins in HMM, whereas BCP is hidden-state free with emphasis on local patterns. Second, four-state CPM model was developed to interpret a single track DNaseI array data while our method was an examination based on multivariate histone modification data. Third, CPM models the DNaseI signal as a continuous mixture of Gaussian at each state, whereas we models histone modifications with a zero-inflated Gaussian mixture due to spatial sparsity of binding events. 

\subsection{Summary and future directions}
In this paper, we have developed a novel multivariate BCP model to partition genome into contiguous blocks based on histone modifications. It could be extended to analyze chip-sequencing data or applied to other studies with partitioning zero-inflated multiple observation tracks as a task. Our model presents a new approach to examining combinatorial histone marks. Not only histone marks are signatures for functional elements (\ca{comprehensive:nature} \cy{comprehensive:nature}, \ca{hmm:naturebiotech} \cy{hmm:naturebiotech}), our results from the \textit{D. \textit{melanogaster}} S2 cell genome suggest that they are also roadmaps for chromatin organization at coarse scales. 

It is worthwhile to further investigate whether BLOCKs and topological domains are substantively different, or if BLOCKs merely re-describes topological domains based on histone modifications. Besides the difference introduced by techniques, data quality and cell types, we believe other two possible reasons for imperfect alignment between BLOCKs and physical domains are: 1) the partition is not saturated based on the current profile of histone modifications; 2) the equal weight assigned to different histone modifications in the partition limit the identification of finer domains (a drawback of all current approaches). 

It has become increasingly clear that functionally related genes are often located next to one another in the linear genome (\ca{linear} \cy{linear}), resembling DNA operon in bacteria  (\ca{dong} \cy{dong}, \ca{operon} \cy{operon}). This proximity is essential for coordinated gene regulation. Genome-wide expression analysis have identified many clusters of co-expressed genes during \textit{\textit{Drosophila}} development (\ca{cluster1} \cy{cluster1}, \ca{cluster2} \cy{cluster2}), such as the \textit{hox} gene clusters (\ca{hox} \cy{hox}). One mechanism for this coordinated regulation is that these genes are organized into a chromatin domain that acts as a regulatory unit by the epigenetic mechanism (\ca{science:domain} \cy{science:domain}, \ca{linear} \cy{linear}). Several such chromatin domains have already been characterized (\ca{science:domain} \cy{science:domain}, \ca{nature:domain1} \cy{nature:domain1}, \ca{nature:domain2} \cy{nature:domain2}, \ca{cell:domain} \cy{cell:domain}). In this study, we illustrated the widespread existence of these chromatin domains as BLOCKs that were identified by combinatorial histone marks. 
 
Last but not least, although we have shown that a substantial portion of BLOCKs can potentially act as regulatory units, this is likely still an underestimate. Firstly, our BLOCKs were identified based on combinatorial patterns of all available 18 histone marks from the S2 epigenome. We do not know in totality how many histone marks are sufficient to saturate the segmentation. It is likely that more markers, including potentially undiscovered ones will be needed to get a complete view of epigenetic landscape. Over 100 histone marks have been discovered yet with a lot of exclusivity and correlation. Future studies addressing relationships among histone marks will give us more insight into this open question. It is also important to develop block identification methods that can accommodate the dependency structure among marks. Secondly, when evaluating expression of genes within an individual BLOCK, we used developmental transcriptome from \textit{Drosophila} tissues other than S2 cells, which only present a weighted average of varying BLOCKs across different cell types within each developmental stage. In reality, each type of cells is likely to have its distinct pattern of BLOCKs. Thirdly, plasticity in chromosomal modifications has been shown in several reports (\ca{heterochromatin:gr} \cy{heterochromatin:gr}, \ca{signature:gr} \cy{signature:gr}, \ca{functional:science} \cy{functional:science}). Thus we would expect BLOCKs are dynamic structures and the percentage of BLOCKs with tendency of co-regulation might be even higher if taking into account this plasticity. This conjecture could be tested when more histone marks data across development stages are available. Fourthly, with incomplete and inaccurate knowledge on gene functions in GO database (as well as others) (\ca{plos:go} \cy{plos:go}), likely many BLOCKs with functional relevance may not stand out just because supporting information doesn't exist yet. Finally, coordinated regulation is a complex process accomplished by miRNA, transcript factors and other regulatory elements with feedback effect on chromatin organization. Further analysis on binding sites of regulatory elements and their interplay with genes within BLOCKs will shed more lights on understanding the underlying mechanism.

\section*{Acknowledgements} We thank the reviewers for their constructive comments and Chao Gao for discussion.

\begin{supplement}
\stitle{Supplementary Figures and Tables} 
\slink[url]{http://www.mengjiechen.com/publication.html}
\sdescription{\\
Figure S1. Number of enriched regions of 46 histone marks and non-histone chromosomal proteins from modENCODE project. \\
Table S1. BLOCKs identified by BCP in S2 cells using posterior probability cutoff 0.75. \\
Table S2. Gene lists in GO enriched BLOCKs in S2 cell. \\
Table S3. BLOCKs with the top 20\% largest deviations in the transcription across 9 different developmental stages. \\
Table S4. BLOCKs with the top 20\% smallest deviations in the transcription across 9 different developmental stages. \\}
\end{supplement}

\bibliography{plos}

\end{document}